\documentclass[aps,pre,twocolumn,showpacs,preprintnumbers,amsmath,superscriptaddress]{revtex4}
\usepackage{amsmath} \usepackage{graphicx}

\begin{document}

\begin{abstract} 
  Stochastic sandpiles self-organize to an absorbing-state critical
  point with scaling behavior different from directed percolation (DP)
  and characterized by the presence of an additional conservation law.
  This is usually called C-DP or Manna universality class. There
  remains, however, an exception to this universality principle: a
  sandpile automaton introduced by Maslov and Zhang, which was claimed
  to be in the DP class despite of the existence of a conservation
  law.  In this paper we show, by means of careful numerical
  simulations as well as by constructing and analyzing a field theory,
  that (contrarily to what previously thought) this sandpile is also
  in the C-DP or Manna class. This confirms the hypothesis of
  universality for stochastic sandpiles and gives rise to a fully
  coherent picture of self-organized criticality in systems with
  conservation. In passing, we obtain a number of results for the C-DP
  class and introduce a new strategy to easily discriminate between DP
  and C-DP scaling.
\end{abstract}

\title{Confirming and extending the hypothesis of universality in
  sandpiles}

\author{Juan A. Bonachela} 
\author{Miguel A.  Mu\~noz}

\affiliation{ Departamento de
  Electromagnetismo y F{\'\i}sica de la Materia and \\ Instituto de
  F{\'\i}sica Te{\'o}rica y Computacional Carlos I, Facultad de
  Ciencias, Univ. de Granada, 18071 Granada, Spain} 


\date{\today}
\maketitle

\section{Introduction}

Aimed at shedding some light on the origin of scale invariance in many
contexts in Nature, different mechanisms for {\it self-organized
  criticality} (SOC) were proposed during the last two decades
following the seminal work by Per Bak and collaborators
\cite{BTW,Dhar,Sandpiles}.  Sandpiles, ricepiles, and earthquake
toy-models become paradigmatic examples capturing the essence of
self-organization to scale-invariant (critical) behavior without
apparently requiring the fine tuning of parameters \cite{GG,SOC,VZ}.

Sandpiles played a central role in the development of this field
\cite{BTW,Dhar,Sandpiles}. They are metaphors of real systems (as
earthquakes, snow avalanches, stick-slip phenomena, etc.) in which
some type of stress or energy is accumulated at some slow timescale
and relaxed in a much faster way.  In sandpiles, grains are slowly
added until eventually they relax if a local instability threshold is
overcome; then, they are transmitted to neighboring sites which, on
their turn, may become unstable and relax, generating avalanches of
activity. Considering open boundaries to allow for energy balance, a
steady state with power-law distributed avalanches is eventually
reached.


In order to rationalize sandpiles in particular and SOC in general and
to understand their critical properties, it was proposed
\cite{DVZ,FES,BJP} to look at them as systems with many absorbing
states \cite{AS,many}.  The underlying idea is that, in the absence of
external driving, sandpile models get eventually trapped into stable
configurations from which they cannot escape, {\it i.e.} absorbing
states \cite{AS}. The concept of {\it fixed energy sandpiles} (FES)
was introduced to make this connection more explicit.  FES sandpiles
share the microscopic rules with their standard (slowly driven and
dissipative) counterparts, but with no driving nor dissipation. In
this way, the total amount of sand or energy becomes a conserved
quantity acting as a control parameter.  Calling $E_c$ the average
density of energy of a standard sandpile in its stationary
(self-organized) critical state, it has been shown that the
corresponding fixed-energy sandpile exhibits a transition from an
active phase to an absorbing one at, precisely, $E_c$, while it is
absorbing (active) below (above) $E_c$. It can be argued that {\it
  slow driving and dissipation acting together at infinitely separated
  time-scales constitute a mechanism able to pin a generic system with
  absorbing states and a conservation law to its critical point}
\cite{GG,VZ,DVZ,FES,BJP,abs_others}.

Using the relation with transitions into absorbing states \cite{interfaces}, a
Langevin equation describing FES stochastic sandpiles was proposed
\cite{BJP,FES} in the spirit of Hohenberg and Halperin \cite{HH}.  The
Langevin equation is similar to the well known directed-percolation (DP)
Langevin equation describing generic systems with absorbing states
\cite{conjecture,AS} but it is coupled linearly to a conserved non-diffusive
energy field, representing the conservation of energy grains in the sandpile
dynamics \cite{FES,BJP}:
\begin{eqnarray}
  \displaystyle{\partial_t \rho (x,t)}& = & 
   a \rho - b \rho^2 + \omega \rho E(x,t) +  \nabla^2 \rho +
  \sigma \sqrt{\rho} \eta(x,t) \nonumber \\ 
  \displaystyle{ \partial_t E (x,t)}& = & D \nabla^2 \rho,
\label{FES}  
\end{eqnarray}
\noindent
where $D$, $a$, $b$ and $\omega$ are constants, $\rho(x,t)$ and $E(x,t)$ are
the activity and the energy field, respectively, and $\eta$ is a zero-mean
Gaussian white noise.  The universality class described by this Langevin
equation, which includes sandpiles as well as all other systems with many
absorbing states and a non-diffusive conserved field, is usually called Manna
or C-DP class \cite{Rossi,FES,BJP,note}.  As an important side note, let us
remark that the original Bak-Tang-Wiesenfeld model, being deterministic, has
many other conservation laws (toppling invariants) and is therefore not
described by the present stochastic theory \cite{deter1,BJP}.

Incidentally, even if it has been clearly established that DP and C-DP
constitute two different universality classes \cite{Romu,Lubeck,DCM}, most of
their universal features (critical exponents, moment ratios, scaling
functions,...) are very similar, making it difficult to discriminate
numerically between both classes in any spatial dimension.

Despite of the fact that the critical behavior of stochastic sandpiles
is accepted to be universal and described by Eq.(\ref{FES}), there are
a few sandpile models which, despite of being conservative, have been
argued to exhibit a different type of critical behavior \cite{MD,MZ},
namely DP, violating the conjecture of universality:

\begin{itemize}
\item the {\it Mohanty-Dhar} (MD) sandpile \cite{MD}, in which grains
  have some probability to remain stable even if they are above the
  instability threshold. It was argued, based on numerical simulations
  and on exact results for an anisotropic version of it, that the
  (isotropic) MD sandpile should exhibit DP behavior,
 
\item the {\it Maslov-Zhang} (MZ) sandpile \cite{MZ}, in which not
  only the grains of unstable sites are redistributed among nearest
  neighbors. Instead, all the grains in the local neighborhood of
  unstable sites are randomly redistributed or ``reshuffled''.
  Various versions of the model (called ``charitable'', ``neutral''
  and ``greedy'') were defined depending on the bias in the local
  redistribution rule (``charitable'' if the central site receives
  less than each n.n., ``neutral'' if all the sites in the
  neighborhood are treated alike, and ``greedy'' otherwise). On the
  basis of Monte Carlo simulations this sandpile (in its neutral
  version) was argued to be DP-like.
\end{itemize}

Aimed at clarifying this puzzling situation, in a recent work we
provided strong numerical and analytical evidence that, contrarily to
what previously thought, the (isotropic) MD sandpile is actually in
the C-DP class \cite{Rapid} (see \cite{MD2} for a discrepant
viewpoint).  To further substantiate our claim, in \cite{Jabo2} we
proposed a strategy to easily discriminate between DP and C-DP
consisting in introducing a wall (either absorbing or reflecting);
systems in the DP and in the C-DP classes behave in (qualitative and
quantitatively) very different ways in the presence of walls,
providing an easy criterion to discriminate between both classes.

Our goal in the present paper is to scrutinize the last remaining piece in the
puzzle, {\it i.e.} the MZ model, by employing extensive numerical simulations
as well as field theoretical considerations.  For the numerics, we take
advantage of the previously introduced discrimination criterion and, also,
present another method to easily discriminate between DP and C-DP, based in
the introduction of anisotropy in one spatial direction.

The analyses presented on what follows, show in a clean-cut way that the MZ
model, in any of the three versions above, is actually in the C-DP class. This
leaves no dangling end in the sandpile universality picture and, as a
byproduct, confirms the general validity of the absorbing-state approach to
rationalize SOC. New results for the C-DP class are also obtained.

\section{The Maslov-Zhang sandpile}

The (neutral) Maslov-Zhang (MZ) sandpile \cite{MZ} is defined as follows:
\begin{enumerate}

\item Driving: an input energy, $\delta E \leq 1$, is added to the
  central (or to a randomly chosen) site, $i$, of a d-dimensional
  lattice, and the site is declared active.

\item Relaxation: energy is locally redistributed (``reshuffled'')
  between the active site and its nearest-neighbors, according to:
\begin{equation}
  E_{i}=  \frac{r_{i}}{\sum_{j=1}^{2d+1}r_{j}}  \displaystyle\sum_{j=1}^{2d+1}E_{j},
\label{reshuffling}
\end{equation}
where $r_{i}$ are uniformly distributed ($r_i \in \left[ 0,1\right] $)
random variables, and the sums are performed over the site $i$ and its
$2d$ nearest neighbors. This rule needs to be slightly modified for
the ``charitable'' and ``greedy'' versions of the model that we will
not explore in detail here.

\item Activation: each of the sites involved in the reshuffling is
  declared active with a probability given by its own energy,
  triggering the generation of avalanches of activity.

\item Avalanches: new active sites are added to a list and relaxed in
  a sequential way.

\item Dissipation: energy arriving at the open borders is removed from
  the system.

\item Avalanches proceed until all activity ceases, and then a new
  external input is added. Eventually a critical stationary state is reached.
\end{enumerate}

Monte Carlo simulations by Maslov and Zhang revealed exponents
compatible with those of directed percolation in two dimensions and
above, while in one dimension some anomalies in the scaling were
reported \cite{MZ}. Later, simulations of the FES counterpart of the
MZ sandpile led to very similar results \cite{FES}.

Before entering a more careful numerical analysis of the MZ sandpile,
we take a detour to construct explicitely a Langevin equation for such
a model. This will allow us to better understand what the main
relevant ingredients of the MZ model are and in what sense it differs
from other C-DP models and from Eq.(\ref{FES}).

\section{A Langevin Equation for the Maslov-Zhang sandpile}

The main difference between the Maslov-Zhang cellular automaton and
other sandpiles in the C-DP universality class is that, while in the
C-DP class as the Manna model, the only energy redistributed by the
dynamics (by topplings) is that accumulated in active sites, in the MZ
dynamics both the energy of active sites as well as that of its
nearest neighbors is redistributed. This leads to a more severe local
redistribution of energy, that we quantify in what follows in terms of
a new set of Langevin equations, first in a phenomenological way and
then by deriving it from the microscopic rules.

\subsection{Phenomenological Langevin equation}
Diffusion of energy occurs in the C-DP class by means of activity
relaxation. At a mesoscopic level, this implies that the rate of
change of the energy density $E(x,t)$ at a given position is
proportional to minus the divergence of a current, $ \partial_t E(x,t
)= - \nabla .  j(x,t)$, where $j(x,t)$ is given by the gradient of the
activity field,
\begin{equation} 
  j(x,t) = - D ~\nabla \rho (x,t)
 \end{equation}
 leading to $ \partial_t E (x,t)= D \nabla^2 \rho(x,t)$ in
 Eq.(\ref{FES}).

 Instead, in the MZ model, changes of energy are controlled by the
 reshuffling rule, {\it i.e.} energy does not need to be at an active
 site (but in its neighborhood) to be redistributed. Hence, at a
 mesoscopic level the current is given by {\it gradients of the energy
   in the presence of non-vanishing activity}. More specifically, $
\partial_t E(x,t )= - \nabla . j(x,t)$ where now
\begin{equation} 
  j(x,t) = - \tilde{D} ~\nabla E (x,t),
 \end{equation}
 and the diffusion $\tilde{D}$ is not a constant but a functional proportional
 to $\rho(x,t)$, $\tilde{D}(\rho(x,t))= D \rho(x,t) $ capturing the
 requirement that local reshuffling of energy only occurs in the presence of
 activity. This enforces the absorbing state condition that dynamics ceases if
 $\rho=0$.  Finally,
\begin{equation}
  \partial_t E(x,t )= D ~\nabla  .  [\rho (x,t) \nabla E(x,t)].
\label{LangevinE}
\end{equation}
 On the other hand, the equation for the activity is not expected to change in
 any relevant way, so one obtains
\begin{eqnarray}
  \displaystyle {\partial_t \rho (x,t)} & = & a \rho - b \rho^2 + \nabla^2 \rho + \omega \rho
  E(x,t) + \sigma \sqrt{\rho} \eta(x,t) \nonumber \\ 
  \displaystyle{ \partial_t E (x,t)} & = & D \nabla . [ \rho(x,t) \nabla E(x,t)]
\label{FES-MZ}  
\end{eqnarray}
representing the MZ dynamics at a mesoscopic level. This is to be compared
with Eq.(\ref{FES}).

\subsection{Microscopic derivation of the Langevin equation}

To gain more confidence on the phenomenological set of equations, Eq.
(\ref{FES-MZ}) we present here an explicit microscopic derivation.  We
consider a parallel version of the MZ model in which all active sites are
relaxed at every time step. To do so, we assume that each site uses a fraction
$1/(2d+1)$ of its total energy for eventual redistributions with each of the
sites in its local neighborhood. At mean-field level, the energy evolves
according to
\begin{equation}
\begin{array}{rl}
  E_{i,t+1}=& \dfrac{1}{2d+1}E_{i,t}\displaystyle\sum_{j=1}^{2d+1}
  \left[ 1- \Theta(\rho_{j,t})\right]  \\
  +&\dfrac{1}{2d+1}\displaystyle\sum_{j=1}^{2d+1}\Theta(\rho_{j,t})
  \sum_{k=1}^{2d+1}\dfrac{E_{k,t}}{2d+1}
\end{array}
\label{Eq1}
\end{equation}
\noindent
where $j$ (respectively $k$) runs over the nearest neighbors of $i$
($j$) and $\Theta$ is the Heaviside step function ($\Theta(z) =0$ if
$z \leq 0$).  For any inactive site in the local neighborhood (i.e.
$\rho_j =0$), the central site does not redistribute the corresponding
fraction of its energy (first term in Eq.(\ref{Eq1})). For any active
site in the local neighborhood (i.e.  $\rho_j > 0$) the central site
receives on average a corresponding fraction of energy (second term).
Eq.(\ref{Eq1}) is a mean-field equation to which fluctuations should
be added to have a more detailed description.  Such fluctuations
appear as a (conserved) noise for the resulting energy equation and
can be easily argued to constitute a higher order, irrelevant,
correction.

Regularizing the Heaviside step function in Eq.(\ref{Eq1}) by means of
a hyperbolic-tangent and expanding it in power series up to first
order in $\rho$ \cite{Rapid}, we obtain
\begin{equation}
\begin{array}{rl}
  E_{i,t+1}=& \dfrac{1}{2d+1}E_{i,t}\displaystyle\sum_{j=1}^{2d+1} \left[ 1-
  \rho_{j,t} \right]  \\ 
 +& \dfrac{1}{2d+1}\displaystyle\sum_{j=1}^{2d+1}
  \rho_{j,t} \sum_{k=1}^{2d+1}\dfrac{E_{k,t}}{2d+1}.
\end{array}
\label{Eq3}
\end{equation}
Introducing the $d$-dimensional discrete Laplacian $\nabla^{2}E_{i}=
\sum_{j=1}^{2d}(E_{j}-E_{i})$ the first term can be rewritten as:
\begin{equation}
\begin{array}{rl}
  \dfrac{E_i}{ 2d+1}\displaystyle\sum_{j=1}^{2d+1}\left[ 1-
    \rho_{j}\right] 
  = & E_{i}-\dfrac{E_i}{2d+1}\displaystyle\sum_{j=1}^{2d+1}\rho_{j} \\
  = &E_{i}-E_{i}\rho_{i}-\dfrac{E_i}{2d+1}\nabla^{2}\rho_{i}.
\end{array}
\label{Eq4}
\end{equation}
\noindent
Similarly, the second term can be expressed as
\begin{equation}
\begin{array}{rl}
\dfrac{1}{(2d+1)^{2}}\displaystyle\sum_{j=1}^{2d+1}\rho_{j}
\left(\displaystyle\sum_{k=1}^{2d+1}E_{k}\right) =&
E_{i}\rho_{i}+\dfrac{1}{(2d+1)}\nabla^{2}\left(E_{i}\rho_{i}\right) \\
+\dfrac{1}{(2d+1)}\rho_{i}\nabla^{2}E_{i}
+&\dfrac{1}{(2d+1)^{2}}\nabla^{2}\left(\rho_{i}\nabla^{2}E_{i}\right).
\end{array}
\label{Eq5}
\end{equation}
\noindent
Putting these two contributions together and reorganizing the discrete
derivatives, one obtains:
\begin{equation}
\begin{array}{ll}
&  E_{i,t+1}= E_{i,t}+\dfrac{2}{2d+1}\rho_{i,t}\nabla^{2}E_{i,t}+
  \dfrac{2}{2d+1}\nabla\rho_{i,t} ~. \nabla E_{i,t} \\  & + ~
  \dfrac{1}{(2d+1)^{2}}\nabla^{2}\left(\rho_{i,t}\nabla^{2}E_{i,t}\right) =\\
&  E_{i,t}+\dfrac{2}{2d+1}\nabla . \left(\rho_{i,t}\nabla E_{i,t}\right)
  + \dfrac{1}{(2d+1)^{2}}\nabla^{2}\left(\rho_{i,t}\nabla^{2}E_{i,t}\right)
\end{array}
\label{Eq6}
\end{equation}
\noindent
which in the continuous time limit becomes
\begin{equation}
  \partial_{t}E_{i}=
  \dfrac{2}{2d+1}\nabla . \left(\rho_{i,t}\nabla E_{i,t}\right)
  +\dfrac{1}{(2d+1)^{2}}\nabla^{2}\left(\rho_{i,t}\nabla^{2}E_{i,t}\right).
  \label{LangevinE2}
\end{equation}
\noindent
The second term in this last equation can be argued to be na\"ively
irrelevant in the renormalization group sense (as it includes higher
order derivatives) and hence dropped out. Identifying $D$ with
$2/(2d+1)$ we recover the phenomenological Eq.(\ref{LangevinE}), as
the leading contribution.

As said above, this is to be compared with the standard Langevin
Eq.(\ref{FES}) for the C-DP class, in which the flux of energy is
given by the gradient of the activity itself.  The following questions
pop up naturally:
Does Eq. (\ref{FES-MZ}) lead to a critical behavior different
  from that of Eq.(\ref{FES})?
Could this be the reason why the MZ sandpile was claimed to
  differ from the C-DP class?
Is this new form of conserved energy dynamics irrelevant at the
  DP fixed point, supporting the MZ sandpile to be DP like?

  To properly answer these questions one should resort to a full
  renormalization group calculation. Given that even for the C-DP
  class this has proven to be a, still not satisfactorily
  accomplished, difficult task \cite{Fred}, we will leave aside such a
  strategy here.  Instead, in section IV we will give an answer to
  these question by means of computational studies of the microscopic
  MZ model as well as of its equivalent Langevin Eq.(\ref{FES-MZ}).

  \subsection{Na\"ive power counting}

  A power counting analysis is not helpful in elucidating the
  relevancy of the energy-diffusion term, Eq.(\ref{LangevinE}).
  Actually, as there is a lineal dependence on the energy field in
  both the r.h.s.  and the l.h.s. of Eq.(\ref{FES}), there is no way
  to extract the energy field na\"ive dimension nor, hence, to make
  any statement about the relevancy of the coupling term $w \rho(x,t)
  E(x,t)$ at the DP fixed point.

  On the other hand, it is easy to cast Eq.(\ref{FES-MZ}) into a
  generating functional following standard procedures, to set the
  basis of a perturbative expansion. However, one soon realizes that
  technical difficulties similar to those encountered for the analysis
  of Eq.(\ref{FES-MZ}) (including the presence of generically singular
  propagators \cite{Fred,FES}) show up, hindering the perturbative
  calculation. It is our believe that some type of non-perturbative
  technique, or non-conventional perturbative expansion, is required
  to elucidate the renormalization group fixed point of this type of
  problems.

  The analytical understanding, at a field theory level of C-DP as
  well as the MZ-Langevin equation remain open challenging tasks.

\section{Monte Carlo simulations of the Maslov-Zhang sandpile}

We have performed extensive Monte Carlo simulations of the MZ sandpile
(and variations of it) and scrutinized its asymptotic (long time and
large system size) properties.  We report on two different types of
numerical experiments.

\begin{itemize}

\item {\it ``SOC''} or avalanche experiments:

  \noindent By iterating slow addition of grains in the sandpile with
  open boundaries, the system self-organizes to a state with average
  energy $E_c$. Then, the avalanche size distribution, $P(s)$ and the
  avalanche time distribution $P_t(t)$, can be estimated and their
  corresponding exponents, $\tau$ and $\tau_t$ \cite{avalanches},
  measured.

\item Absorbing state experiments:

  \noindent 
  At the stationary state, we perform spreading experiments from a
  localized seed, and we measure: i) the mean quadratic distance to
  the initial seed $R^{2 } \sim t^{z_{spr}}$ in active runs, ii) the
  average number of active sites as a function of time, $N(t)\sim
  t^{\eta}$ and iii) the surviving probability up to time $t$, $P_{s}
  \sim t^{-\delta}$.  We also study the decay at criticality of a
  homogeneous initial activity ($\rho(t) \sim t^{-\theta}$) in the
  fixed energy case \cite{AS,avalanches}.
\end{itemize}

Scaling laws relating avalanche to spreading exponents where described
systematically in \cite{avalanches}; two of them are:
$\tau~=(1+\eta+2\delta)/(1+\eta+\delta)$, $\tau_{t}~=1+\delta$.  We
measure the exponents independently and use scaling laws as a check
for consistency.

We simulate the MZ automaton in one-dimensional lattices up to size
$L=2^{15}$.  The stationary critical energy density is
$E_{c}=0.4928(2)$ and, contrarily to what reported in \cite{MZ}, we do
observe clean scaling at criticality, even if it emerges only after
significatively long transients (results not shown) justifying why
smaller scale simulations can lead to erroneous conclusions.

The resulting critical exponents are gathered together in Table
\ref{table_I}. They are closer in all cases to the C-DP values than to
DP ones. The exponents measured explicitely in \cite{MZ} are $\alpha$
($\int dt N(t)/P_s(t) \sim t^\alpha$) and the fractal dimension $D_f$, which
take also very similar values in both cases ($\alpha = 1.47(1)$, $D_f=
2.32(1)$ for DP, $\alpha= 1.52(1)$, $D_f = 2.36(1)$ for C-DP).  Still,
given the similarity between the numerical values in both classes, it
is not save to extract a definitive conclusion from these results.
Higher numerical precision would be required to produce fully
convincing evidence.
\begin{table} [ht!]
\begin{centering}
\begin{tabular}{|l|c|c|c|c|c|c|}
\hline
${\bf d=1}$&$\eta$&$\delta$&$\tau$&$\tau_{t}$&$z_{spr}$&$\theta$ \\
\hline
\hline
DP& $0.313(1)$& $0.159(1)$&$1.108(1)$ &$1.159(1)$&$1.265(1)$&$0.159(1)$ \\
\hline
\hline
C-DP& $0.350(5)$& $0.170(5)$&$1.11(2)$  &$1.17(2)$&$1.39(1)$&$0.125(1)$ \\
\hline
\hline
MZ & $ 0.32(5) $& $0.20(5)$&$1.13(5)$ &$1.20(5)$&$1.40(5)$&$0.13(1)$\\
\hline
\end{tabular}
\caption{\footnotesize{Critical exponents for DP, C-DP,
    and the MZ sandpile in one dimension. DP and C-DP values taken
    from \cite{avalanches,Lubeck,Jabo2}.}}
\label{table_I}
\end{centering}
\end{table}
As said above, the numerical values of DP and C-DP exponents are
closer and closer as the dimensionality is increased (they coincide
above $d_c=4$).  Therefore, performing numerical simulations, as the
ones above, to discriminate between both classes by using larger and
larger times and system sizes in $d \geq 2$ , is not a clever idea.
Instead, it is advisable to use/devise more effective numerical
strategies to discriminate between DP and C-DP in a simple, efficient,
and numerically inexpensive way. For this we use:
\begin{itemize}
\item the method devised in \cite{Jabo2} consisting in analyzing how
  the system responds to the presence of a wall, or

\item a new strategy, which exploits the fact that systems in these
  two classes react in remarkably different ways to the introduction
  of anisotropy in space.
\end{itemize}

Both of these strategies allow to obtain clean-cut results, as shown
in the forthcoming two subsections.

\subsection{Boundary Driven Experiments}
\label{how_to_I}

The influence of walls in systems in the DP class has been profusely
analyzed in the literature \cite{DP_wall}.  In particular, it is well
known that, if spreading (and SOC) experiments are performed nearby a
wall, the surviving probability is significatively affected, and
avalanche and spreading exponents change in a non-trivial way with
respect to their bulk counterparts \cite{DP_wall}.  It is also known
that in the DP class, both reflecting and absorbing walls lead to a
common type of universal ``surface critical behavior'', that we call
surface directed percolation (SDP), characterized (in one-dimension)
by the exponents shown in Table II.

In contrast, the effect of walls in C-DP systems has been studied only
recently \cite{Jabo2}. Contrarily to the DP case, absorbing and
reflecting walls induce different types of surface critical behavior.
As illustrated in Table II, all spreading and avalanche exponents take
distinct values for an absorbing and for a reflecting wall.
Furthermore, the numerical differences between the exponents for
either type of wall with respect to their corresponding SDP
counterparts are very large, allowing for easy numerical
discrimination \cite{Jabo2}.  Finally, in the C-DP class, the
exponents in the presence of a reflecting wall coincide with their
bulk counterparts \cite{Jabo2}.  These features imply that, by
introducing a wall in a given system with absorbing states, it becomes
straightforward to distinguish if it is in the DP or in the C-DP
class, with moderate computational cost.

Following this strategy, we simulated the one-dimensional MZ sandpile,
as defined above, in the presence of both reflecting and absorbing
walls. In both cases a wall is introduced at the origin (position
$i=0$), and the sandpile is studied in the positive half lattice.  In
the {\it reflecting} case, the energy that should go after
reshuffling, to the leftmost site, at $i=0$, (whose energy is fixed to
zero) is instead added to its closest nearest neighbor to the right,
$i=1$.  On the other hand, the {\it absorbing} condition is imposed by
fixing the energy of the leftmost site to zero after every iteration
of the microscopic sandpile rules, i.e. by removing from the system at
every iteration all the energy received by the leftmost site.

\begin{figure}
  \includegraphics[width=70mm]{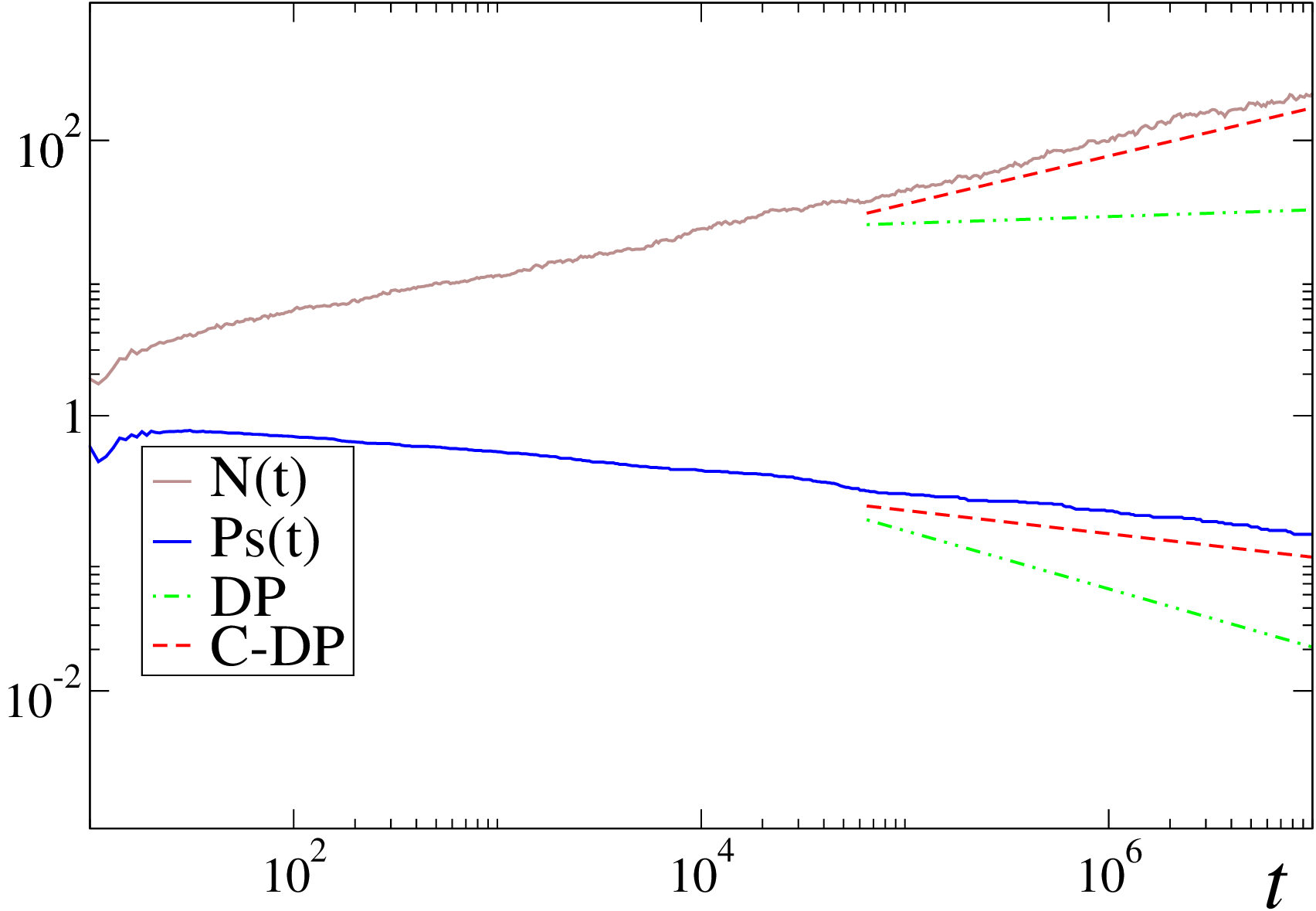}
\includegraphics[width=70mm]{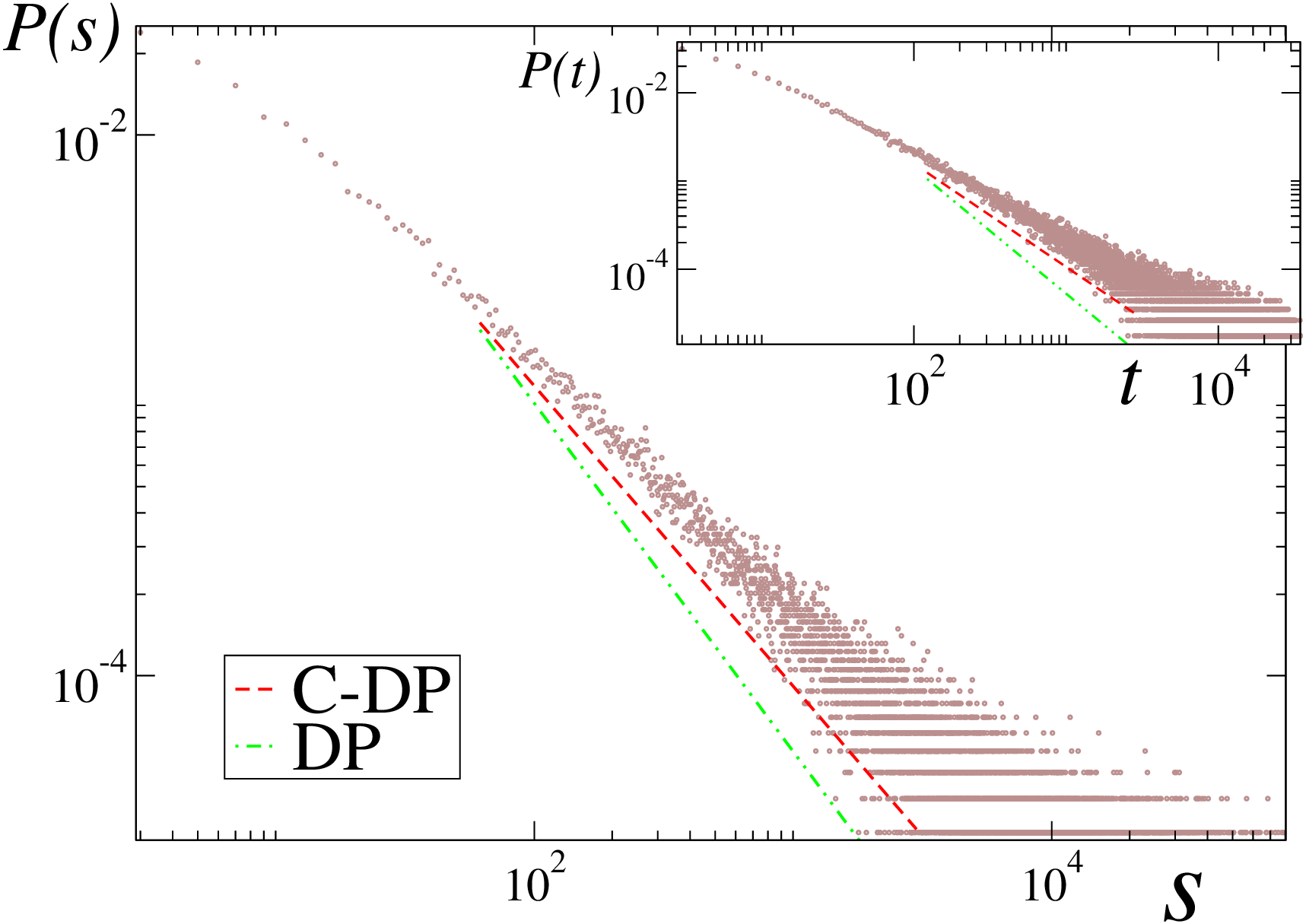}
\caption{(Color online) Avalanche exponents for the one-dimensional MZ
  sandpile in the presence of a {\bf reflecting wall}.  Up: spreading
  experiments; $N(t)$ is the total number of active sites and $P_s(t)$
  is the surviving probability; system size $L=2^{15}$ number of runs
  $500$ . Down: avalanche size (main plot) and time (inset)
  distributions; system size $L=2^{12}$ number of runs $10^5$ . Green
  (red) lines mark DP (C-DP) scaling (see numerical values in Table
  \ref{table_II}). }
\label{reflecting}
\end{figure}
\begin{figure}
\includegraphics[width=70mm]{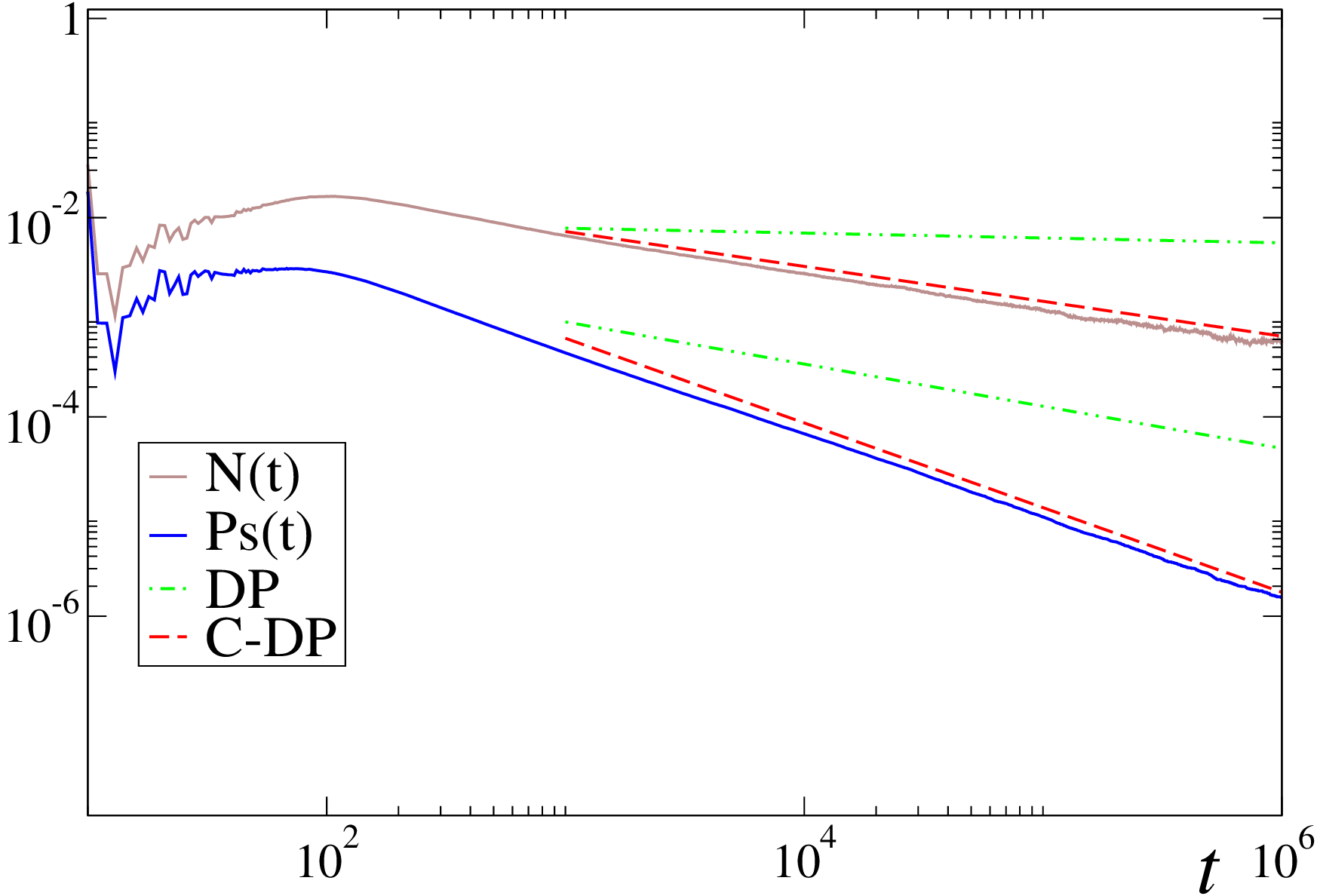}
\includegraphics[width=70mm]{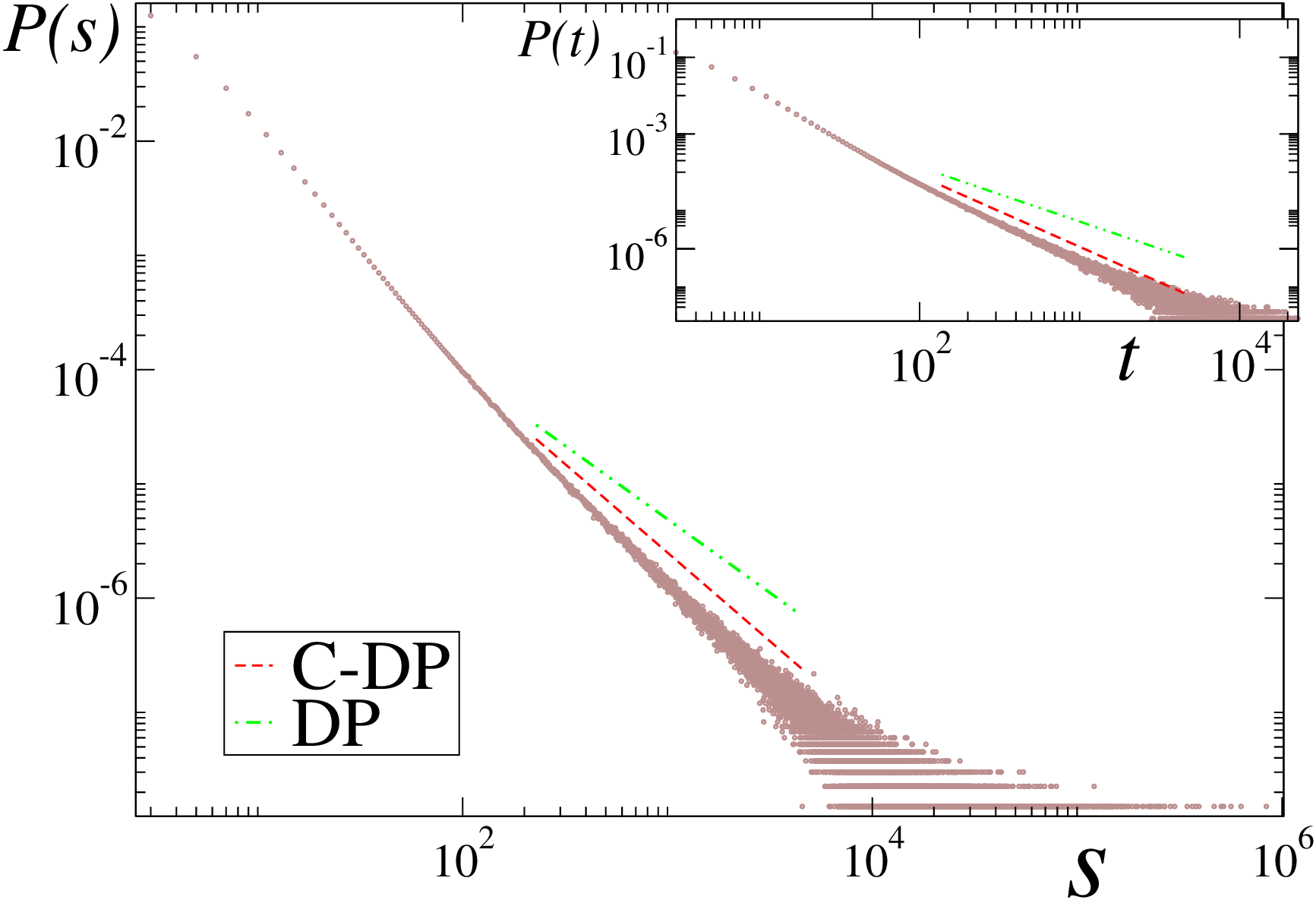}
\caption{(Color online) As in Fig. \ref{reflecting} but for an {\bf absorbing wall}
  (numerical values summarized in Table \ref{table_II}). System size
  $L=2^{15}$ and $8 \cdot 10^7$ runs.}
\label{absorbing}
\end{figure}
\begin{table} [ht!]
  \begin{centering}
\begin{tabular}{|l|c|c|c|c|c|}
  \hline
 $ {\bf d=1}$  &$\eta$&$\delta$&$\tau$&$\tau_{t}$&$z_{spr}$\\
  \hline
  \hline
  DP$_{ref}$&$0.046(2)$&$0.425(2)$&$1.25(3)$&$1.43(3)$&$1.257(2)$\\
  DP$_{abs}$&$0.045(2)$&$0.426(2)$&$1.28(3)$&$1.426(2)$&$1.276(2)$\\
  \hline
  \hline
  C-DP$_{ref}$&$0.35(1)$ &$0.16(1)$&$1.11(2)$&$1.15(2)$&$1.41(1)$\\
  C-DP$_{abs}$&$-0.33(2)$&$0.85(2)$&$1.56(2)$&$1.81(2)$&$1.43(2)$\\
  \hline
  \hline
  MZ$_{ref}$ &$0.37(5)$ &$0.15(5)$&$1.09(5)$&$1.14(5)$&$1.30(5)$\\
MZ$_{abs}$&$-0.36(5)$&$0.84(5)$&$1.57(5)$&$1.84(5)$&$1.25(5)$\\
  \hline
\end{tabular}
\caption{\footnotesize{One-dimensional critical exponents for DP and
    C-DP without walls \cite{avalanches,Lubeck} and in the presence of
    absorbing and reflecting walls \cite{Jabo2}. Values in rows $1$
    (DP$_{ref}$) and $2$ (DP$_{abs}$) coincide within error-bars.
    Note also, that values in row $3$ ( C-DP$_{ref}$ ) coincide with
    those for C-DP (table \ref{table_I}).  Results for the MZ sandpile
    in the presence of reflecting/absorbing walls are reported in the
    last two rows. 
}}
  \label{table_II}
\end{centering}
\end{table}
Fig.(\ref{reflecting}) and Fig.(\ref{absorbing}) show the results of
simulations performed in lattices of system size $L=2^{15}$, averaging
over up to $8 \cdot 10^7$ runs.  The corresponding exponents are
summarized in Table II. All of them coincide within numerical accuracy
with the expected values for the C-DP class in the presence of
reflecting or absorbing walls respectively, and differ significatively
from those of SDP.  For example, in the presence of a reflecting
(respectively, absorbing) wall, the measured value of $\eta$ is $
0.37$ ($-0.37$) in good agreement with the C-DP expectation, $\eta
=0.35$ ($-0.33$) and in blatant disagreement with the corresponding DP
value, $\eta \approx 0.046$ ($\approx 0.045$), which is one order of
magnitude smaller (and of opposite sign in the case of an absorbing
wall).  Similar large differences are measured for all the exponents
(see table II). Note also that, as is the case in the C-DP class
\cite{Jabo2}, the exponents in the presence of a reflecting wall
coincide within error-bars with their bulk counterparts.  In
conclusion, studying the influence of walls we conclude that the
one-dimensional MZ sandpile exhibits C-DP scaling.


\subsection{Anisotropic Experiments}
\label{how_to_II}

It is well known that systems in the DP class are invariant under
Galilean transformations: if particles have a tendency to move
anisotropically in one preferred spatial direction, that does not
alter the critical properties \cite{AS}. The presence of any degree of
anisotropy in DP-like systems is an irrelevant trait, or in other
words, {\it anisotropic DP (A-DP) is just DP}.

The role of anisotropy in sandpiles has also been profusely studied
after the pioneering exact solution by Dhar and Ramaswamy \cite{DR} of
the totally anisotropic or ``directed'' counterpart of the
Bak-Tang-Wiesenfeld sandpile.  Anisotropic stochastic sandpiles have
also been studied using general principles \cite{ani_general} and
through interfacial representations \cite{ani_inter}.  The conclusion
is that all anisotropic sandpiles, as long as they are stochastic
\cite{deterministic}, belong to the same universality class, that we
call {\it anisotropic C-DP} (A-C-DP) \cite{ani_PSV}.  The critical
exponents of models in this class, where first measured numerically
\cite{ani_PSV}, and then exactly calculated in any dimension
\cite{ani_solution} (see table \ref{table_III} and table
\ref{table_IIIb}).
\begin{table} [ht!]
  \begin{centering}
\begin{tabular}{|l|c|c|c|c|c|}
  \hline ${\bf d=1}$
  &$\eta$&$\delta$&$\tau$&$\tau_{t}$&$z_{spr}$\\ \hline
  \hline
  DP &$0.33(2)$&$0.14(2)$&$1.09(3)$&$1.14(3)$&$2.00(2)$\\
  \hline 
  \hline
  A-C-DP &$0$ &$1/2$ &$4/3$ & $3/2$ & $2$  \\
  \hline 
  \hline
  A-MZ&$-0.02(3)$&$0.51(3)$&$1.35(5)$&$1.48(5)$&$1.98(3)$\\
  \hline
\end{tabular}
\caption{\footnotesize{One dimensional critical exponents for DP
    \cite{avalanches,Lubeck}, C-DP with a preferred direction
    (analytical results from \cite{ani_solution}), and the anisotropic
    MZ model.}}  \label{table_III} \end{centering} \end{table}
\begin{table} [ht!]  \begin{centering}
    \begin{tabular}{|l|c|c|c|c|c|} \hline ${\bf d=2}$
      &$\eta$&$\delta$&$\tau$&$\tau_{t}$&$z_{spr}$\\
      \hline \hline
      DP&$0.230(1)$&$0.451(1)$&$1.268(1)$&$1.451(1)$&$1.13(2)$\\
      \hline \hline
      A-C-DP&$0$&$3/4$&$10/7$&$7/4$&$2$\\
      \hline \hline 
      A-MZ&$-0.07(5)$&$0.75(5)$&$1.49(5)$&$1.70(5)$&$1.93(5)$\\
      \hline \end{tabular} \caption{\footnotesize{Two-dimensional
        critical exponents for anisotropic models: A-DP (i.e. DP),
        A-C-DP (analytical results), and anisotropic-MZ.}}  
    \label{table_IIIb} 
\end{centering}
\end{table} 

The strategy to be used is straightforward: take the MZ
sandpile model and switch on anisotropy; if the isotropic model was in
the DP class, anisotropy should be an irrelevant ingredient and the
anisotropic counterpart should also be DP like. If, instead, the
isotropic model is in the C-DP class, then anisotropy is a relevant
ingredient and critical exponent change from C-DP to A-C-DP values.

The simplest way to define an anisotropic MZ (A-MZ) model is by fixing
one of the $r_j$ in Eq.(\ref{reshuffling}), say the one to the right,
to its maximum possible value, $r_j=1$, and letting the others $r_j$
to take randomly distributed values in $\left[0,1\right]$. This
generates an overall energy flow towards the preferred direction (to
the right, in this case). Anisotropy can be introduced in other ways,
including full-anisotropy or directness, but this does not affect our
conclusions in any significant way.

\begin{figure}
\includegraphics[width=70mm]{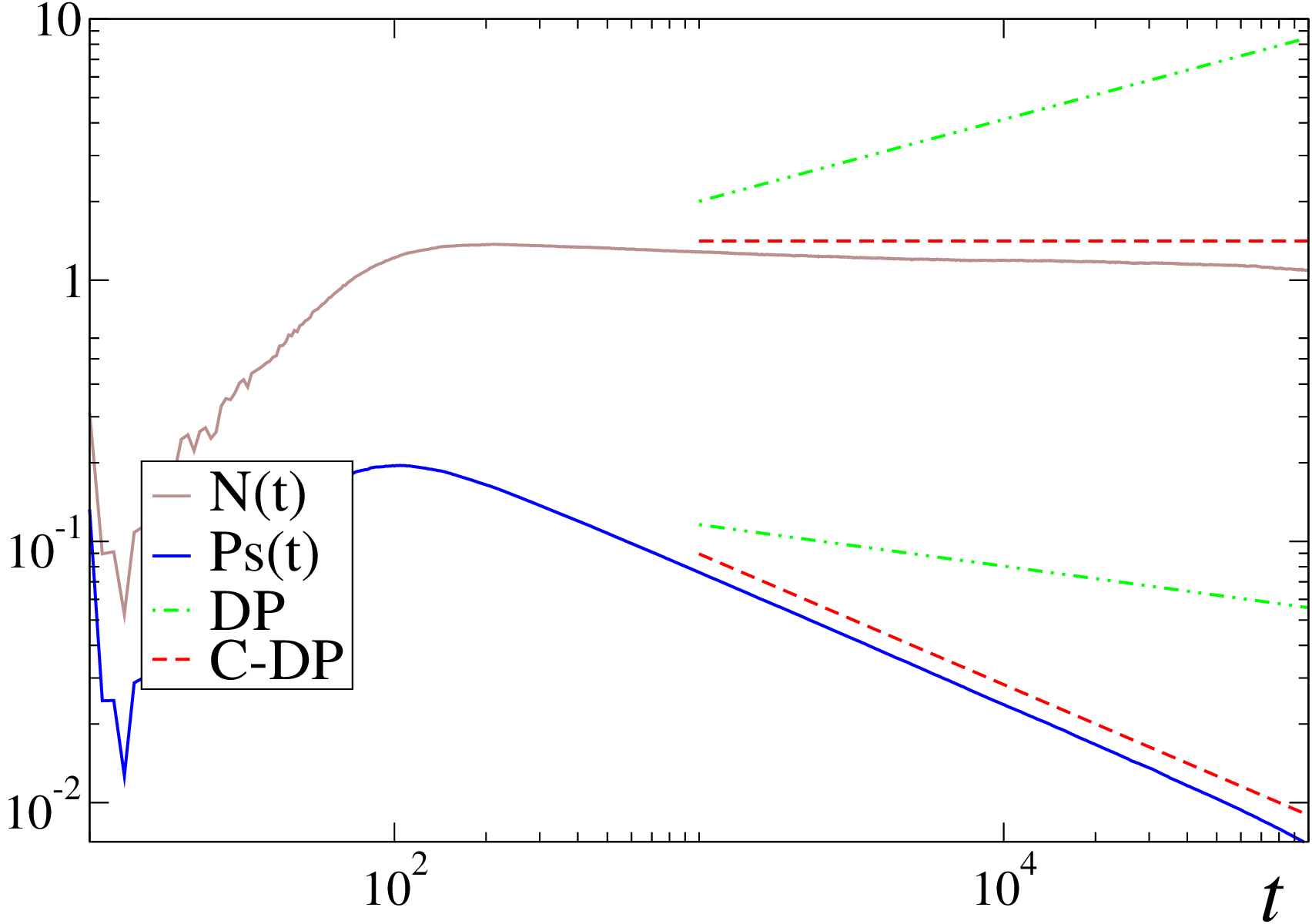}
\includegraphics[width=70mm]{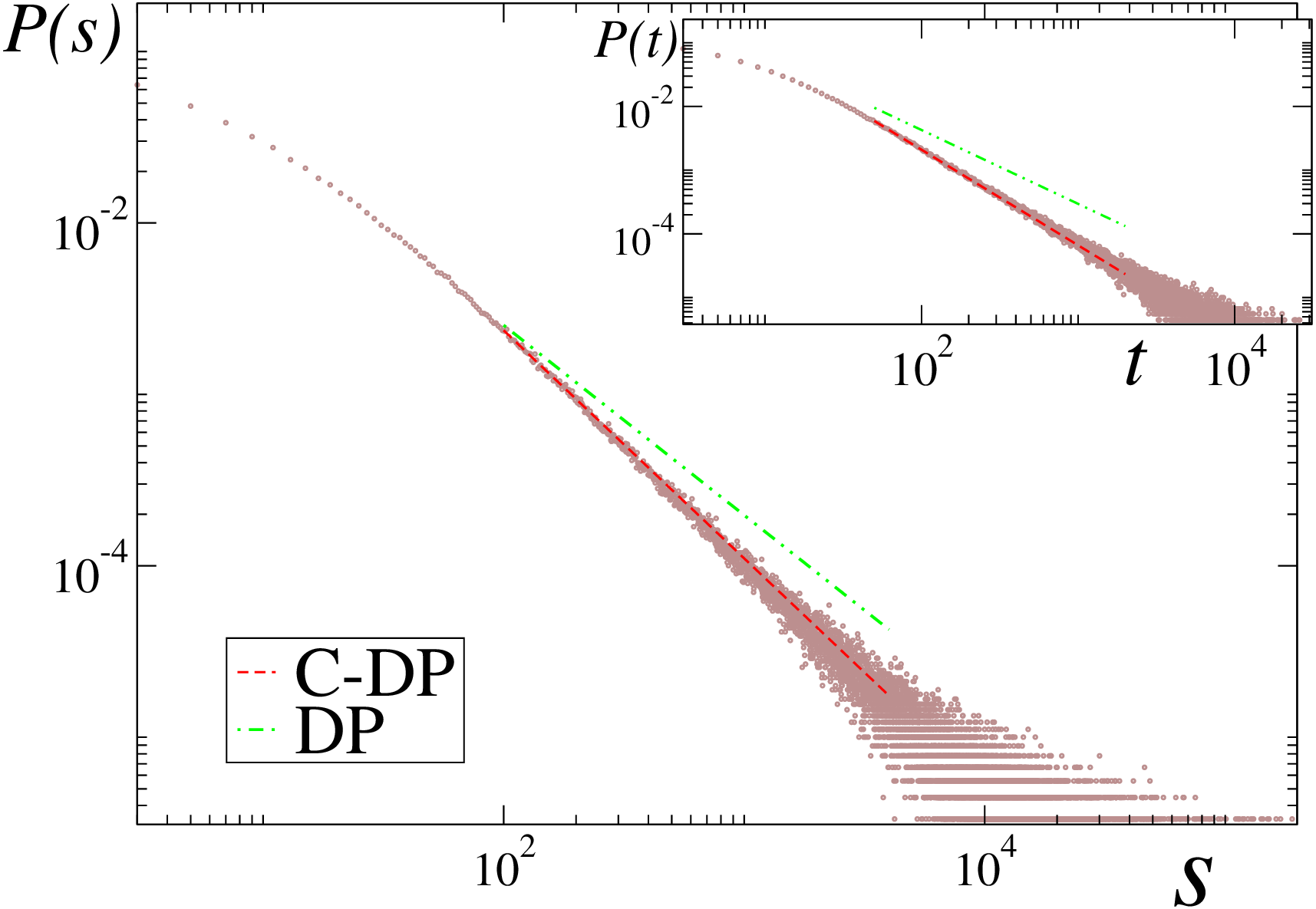}
\caption{(Color online) Avalanche exponents for the {\bf anisotropic} MZ sandpile
  model in one dimension, averaged over $8 \cdot 10^6$ runs (system
  size $L=2^{18}$).  Up: spreading experiments (see Table
  \ref{table_III}).  Down: avalanche size (main plot) and time
  (inset) distributions. Green lines mark DP scaling while red ones
  correspond to C-DP scaling.}
\label{anisotropic}
\end{figure}

Fig. \ref{anisotropic} and table \ref{table_III} show our main results
for the one-dimensional MZ model with anisotropy. Both avalanche and
spreading exponents are very different from their isotropic
counterparts. They also differ notoriously from DP values, but
coincide within error-bars with the expected values for the A-C-DP
class.
The same conclusion holds in two dimensions (see Table
\ref{table_IIIb}). In this way, as the anisotropic MZ model belongs to
A-C-DP class the original, isotropic, MZ sandpile model can be safely
concluded to be in the C-DP universality class, confirming the result
above.

\section{Numerical integration of the MZ Langevin equation}
\label{lang_sim}

In this section we verify that Langevin Eq.(\ref{FES-MZ}) is a sound
description of the MZ model and that, despite of its different form,
it behaves asymptotically as Eq.(\ref{FES}). For that we perform
numerical analysis (again, both SOC and absorbing state experiments)
using Eq.(\ref{FES-MZ}).  A direct integration of Eq.(\ref{FES-MZ}) in
one-dimension, using the recently introduced integration scheme for
Langevin equations with square-root noise \cite{DCM}, produces the
exponents reported in the first row of Table \ref{table_IV} (plots not
shown).  All of them are compatible with those of the microscopic MZ
model and the C-DP class (see Table \ref{table_I}).
\begin{table} [ht!]
\begin{centering}
\begin{tabular}{|l|c|c|c|c|c|c|}
  \hline
  ${\bf d=1}$&$\eta$&$\delta$&$\tau$&$\tau_{t}$&$z_{spr}$&$\theta$\\
  \hline
  \hline
  Eq.(\ref{FES-MZ})&$0.28(5)$&$0.21(5)$&$1.14(5)$&$1.21(5)$&$1.25(5)$&$0.14(2)$\\
  \hline
  \hline
  Eq.(\ref{FES-MZ})$_{ref}$&$0.36(5)$&$0.18(5)$&$1.12(5)$&$1.18(5)$&$1.29(5)$&$0.11(5)$\\
  \hline
  \hline
  Eq.(\ref{FES-MZ})$_{abs}$&$-0.39(5)$&$0.85(5)$&$1.58(5)$&$1.85(5)$&$1.22(5)$&$0.15(5)$\\
  \hline
  \hline
  Eq.(\ref{FES-MZ})$_{anis}$&$-0.01(2)$&$0.50(2)$&$1.39(5)$&$1.64(5)$&$1.98(2)$&$0.51(2)$\\
  \hline
\end{tabular}
\caption{\footnotesize{Critical exponents for Eq.(\ref{FES-MZ}), with
    a reflective wall, Eq.(\ref{FES-MZ}) with an absorbing wall, and
    Eq.(\ref{FES-MZ}) with an anisotropic term.}}
\label{table_IV}
\end{centering}
\end{table}
Changing the boundary conditions during the integration we implement
the reflecting or the absorbing wall. For the former, we impose
$\rho(-x,t)=\rho(x,t), ~~ E(-x,t)=E(x,t),$
\noindent
while for the absorbing walls
$\rho(x \leq 0,t)=0, ~~ E(x \leq 0,t)=0.$
The measured exponents, performing avalanche and spreading experiments
nearby a reflecting (absorbing) wall at $0$ (results not shown) are
summarized in the second and third row of Table \ref{table_IV}.
Again, the exponents coincide within error-bars with their
corresponding C-DP counterparts and exclude DP scaling (see Table
\ref{table_II}).

Finally, we have studied an anisotropic version of the equations by
introducing a term proportional to $\nabla \rho(x,t)$ into both, the
activity and the energy equations in Eq.(\ref{FES-MZ}), obtaining
again excellent agreement with the one-dimensional C-DP values (Table
\ref{table_III}).

In summary, we have integrated numerically Eq.(\ref{FES-MZ}), and
implemented the necessary modifications (i.e. include boundaries or
anisotropy) to perform the tests described in the previous section.
The obtained results are in excellent agreement with those for the
microscopic model, confirming that (i) the Langevin equation derived
in section II is representative of MZ model, and that (ii) the MZ
model is in the C-DP class.

\section{Conclusion and Discussion}

We have shed some light on the picture of universality in stochastic
sandpiles, by confirming that, indeed, they all share the same
universal critical behavior. As hypothesized some years ago, their
critical features are captured by the set of Langevin Eq.(\ref{FES}),
C-DP, describing in a minimal way the phase transition into a multiply
degenerated absorbing state in the presence of a non-diffusive
conserved field.

We have shown that the Maslov-Zhang sandpile, believed before to
exhibit a different type of scaling (directed percolation like), is
actually in the C-DP class, in agreement with the universality
hypothesis. To reach this conclusion we have performed large scale
simulations and introduced new numerical strategies to easily
discriminate between DP and C-DP. In particular, we have benefited
from the fact that the, otherwise very similar, DP and C-DP classes
behave in radically different ways both in the presence of walls and
when anisotropy is switched on. 

We have also derived, in two different ways, an alternative set of
Langevin equations, Eq.(\ref{FES-MZ}), describing the Maslov-Zhang
sandpile. This new set of equations is characterized by a different
form of local energy diffusion (the corresponding current is
proportional to energy gradients and not to activity gradients as is
the case in Eq.(\ref{FES})). By direct integration of the stochastic
differential set of Eq.(\ref{FES-MZ}), we have shown that it describes
the same universality class as Eq.(\ref{FES}), i.e. C-DP, despite of
the formal differences in their respective equations for the conserved
field, hence leading to a coherent global picture for the universality
of sandpiles.  This result actually enlarges the C-DP universality
class, allowing to embrace also different types of energy relaxation
or redistribution dynamics, which includes also ``charitable''
versions of the MZ model. Instead, the ``greedy'' version,
characterized by ``anti-diffusion'' (i.e. energy accumulates on active
sites) is expected to be highly anomalous.

Our analyses have several general implications for the C-DP
universality class:


 i) Reflecting walls are not a relevant perturbation in this class:
  avalanche and spreading exponents measured in the vicinity of a
  reflecting wall coincide with their corresponding bulk counterparts.
  The underlying reason for this remains to be well understood.

 ii)  Absorbing walls are relevant ingredients and affect the
  corresponding surface critical behavior. In particular, avalanches
  and spreading experiments performed nearby the wall are
  characterized by exponents that differ from their bulk counterparts.

iii) Anisotropy in space is also a relevant ingredient. The
  corresponding critical behavior is described by the set of Langevin
  Eqs.(\ref{FES}) (or, equivalently Eq.(\ref{FES-MZ})) with an extra
  term $\nabla \rho(x,t)$ in both equations.  Contrarily to the
  isotropic case, the critical exponents in the anisotropic class are
  known exactly in any dimension. The results coincide with those of
  anisotropic interfaces in random media, confirming once again the
  equivalence between the absorbing-state and the interface pictures
  for SOC sandpiles \cite{interfaces}.

It would be highly desirable to have a working renormalization group
calculation allowing to put all the results discussed here under a
solid analytical ground.

\begin{acknowledgments} We are indebted to our colleagues Hugues
  Chat\'e and Ivan Dornic who participated in the early stages of this
  work.  Support from the Spanish MEyC-FEDER, project FIS2005-00791,
  and from Junta de Andaluc{\'\i}a as group FQM-165 is acknowledged.
\end{acknowledgments}


\begin{thebibliography}{99}
\bibitem{note} Eq.(\ref{FES}) has also been derived rigorously by
  employing standard Fock-space formalism techniques for a
  reaction-diffusion model in this class \cite{Romu}.
\end{thebibliography}
\end{document}